\documentclass[preprint,showpacs,floats,floatfix,nofootinbib]{revtex4}

\usepackage{amsbsy}
\usepackage{amssymb}
\usepackage{amsmath}
\input{epsf}
\usepackage{graphicx}
\usepackage{amssymb}
\usepackage{bm}
\def\spose#1{\hbox to 0pt{#1\hss}}

\def\lta{\mathrel{\spose{\lower 3pt\hbox{$\mathchar"218$}}
     \raise 2.0pt\hbox{$\mathchar"13C$}}}
\def\gta{\mathrel{\spose{\lower 3pt\hbox{$\mathchar"218$}}
     \raise 2.0pt\hbox{$\mathchar"13E$}}}

\newcommand{\ie}{\textsl{i.e.~}}
\newcommand{\cf}{\textsl{cf.~}}
\newcommand{\eg}{\textsl{e.g.~}}

\usepackage{color}

\def\beq{\begin{equation}}
\def\eeq{\end{equation}}
\def\bea{\begin{eqnarray}}
\def\eea{\end{eqnarray}}

\def\x{{\rm x}}
\def\y{{\rm y}}

\def\p{{\rm p}}
\def\e{{\rm e}}
\def\s{{\rm s}}

\def\fint{\int_\mathcal{M} {\rm d}^4 x \sqrt{- g}}

\def\jacx{{}^\x{\cal J}^{A B C}_{a b c}}
\def\jacxbcd{{}^\x{\cal J}^{A B C}_{b c d}}

\def\jacybcd{{}^\y{\cal J}^{A B C}_{b c d}}

\def\Bx{{\cal B}^\x}

\def\Axy{{\cal A}^{\x \y}}

\def\nx{n_\x}

\def\lefx{{\cal L}_{\xi_\x}}
\def\lefy{{\cal L}_{\xi_\y}}

\def\RXYa{R^{\x \y}_a}
\def\RXYb{R^{\x \y}_b}
\def\RXYbara{\bar{R}^{\x \y}_a}
\def\RYXbara{\bar{R}^{\y \x}_a}
\def\RYXa{R^{\y \x}_a}

\def\gtotxa{G_a^\x}
\def\GXYa{G^{\x \y}_a}
\def\GYXa{G^{\y \x}_a}

\def\jacx{{}^\x\mathcal{J}^{A B C}_{a b c}}

\def\jacxbcd{{}^\x\mathcal{J}^{A B C}_{b c d}}

\def\gx{\gamma_\x}
\def\Gx{\Gamma_\x}

\def\rtotxa{R_a^\x}

\def\rtotxbara{\bar{R}_a^\x}

\def\rtotpa{R_a^\p}
\def\rtotea{R_a^\e}
\def\rtotsa{R_a^\s}

\def\hxab{g^{A B}_\x}
\def\hxac{g^{A C}_\x}
\def\hxad{g^{A D}_\x}

\def\hxbe{g^{B E}_\x}
\def\hxca{g^{C A}_\x}
\def\hxcd{g^{C D}_\x}
\def\hxcf{g^{C F}_\x}
\def\hxde{g^{D E}_\x}
\def\hxef{g^{E F}_\x}
\def\hxfg{g^{F G}_\x}

\def\hxinvab{g^\x_{A B}}

\def\hxinvad{g^\x_{A D}}

\def\hxinvbc{g^\x_{B C}}

\def\hxinvbe{g^\x_{B E}}
\def\hxinvcb{g^\x_{C B}}

\def\hxinvcf{g^\x_{C F}}
\def\hxinvad{g^\x_{A D}}

\def\hxyab{g^{A B}_{\x \y}}
\def\hxyac{g^{A C}_{\x \y}}
\def\hxyad{g^{A D}_{\x \y}}

\def\hxybe{g^{B E}_{\x \y}}

\def\hxycd{g^{C D}_{\x \y}}

\def\hxycf{g^{C F}_{\x \y}}

\def\hxyinvab{g_{A B}^{\x \y}}

\def\hxyinvad{g_{A D}^{\x \y}}
\def\hxyinvbc{g_{B C}^{\x \y}}

\def\hxyinvbe{g_{B E}^{\x \y}}

\def\hxyinvcf{g_{C F}^{\x \y}}

\def\hyxba{g^{B A}_{\y \x}}

\def\hyxca{g^{C A}_{\y \x}}
\def\hyxcd{g^{C D}_{\y \x}}

\def\hyxef{g^{E F}_{\y \x}}

\def\hyxinvab{g_{A B}^{\y \x}}
\def\hyxinvba{g_{B A}^{\y \x}}
\def\hyxinvcb{g_{C B}^{\y \x}}

\begin{document}
\title{A variational approach to resistive relativistic plasmas}
\author{N. Andersson$^1$,  G.L. Comer$^2$ and I.Hawke$^1$  }
\affiliation{
$^1$ Mathematical Sciences and STAG Research Centre, University of Southampton,
Southampton SO17 1BJ, United Kingdom\\
$^2$ Department of Physics, Saint Louis University, St. Louis, MO, 63156-0907, USA}

\date{\today}

\begin{abstract}
We develop an action principle to construct the field equations for a multi-fluid system 
containing charge-neutral fluids, plasmas, and dissipation (via resistive interactions), by 
combining the standard, Maxwell action and minimal coupling of the electromagnetic field 
with a recently developed action for relativistic dissipative fluids. \ We use a pull-back 
formalism from spacetime to abstract matter spaces to build unconstrained variations for 
both the charge-neutral fluids and currents making up the plasmas. \ Using basic linear 
algebra techniques, we show that a general ``relabeling'' invariance exists for the abstract 
matter spaces. \ With the field equations in place, a phenomenological model for the 
resistivity is developed, using as constraints charge conservation and the Second Law of 
Thermodynamics. \ A minimal model for a system of electrons, protons, and heat is 
developed using the Onsager procedure for incorporating dissipation. 
\end{abstract}


\maketitle

\section{Introduction}
\label{intro}

Relativistic fluid dynamics is a well developed area of research, with exciting applications 
ranging from astrophysics to high-energy collider physics (see Andersson and Comer 
\cite{andersson07:_livrev}). \ These applications become more complex and involved as our 
computational technology advances. \ In astrophysics, the state-of-the-art is represented by 
neutron-star simulations (or work on supernova core collapse) including multi-dimensional 
neutrino transport \cite{SNreview,snneut} and compact  mergers  of magnetised binary stars including 
resistive effects \cite{baiottirezzollareview,kikibinary}. \ Meanwhile, the high-energy 
physics problem has inspired the first simulations of second-order, causal, dissipative 
models, building on the classic formulation of Israel and Stewart 
\cite{Israel79:_kintheo1,Israel79:_kintheo2}. \ In parallel, there have been formal 
developments of the theory (including many relevant efforts in the string-theory inspired 
area of holography, see for example, \cite{2009CQGra..26v4003R}). 

When it comes to classical (general-) relativistic fluid dynamics, the most interesting 
developments involve the consideration of multi-fluid systems, \eg issues arising when 
components become superfluid, when heat flows and when the electromagnetic charge 
current is treated as a dynamical variable \cite{andersson07:_livrev, heat1,heat2,namhd}. \ These advances allow us to consider a 
wide range of relevant phenomena, but the general theory is incomplete in two important 
respects. \ First of all, we need to be able to consider dissipation (for which a plethora of 
mechanisms may operate in a multifluid system). \ Secondly, we need to couple the 
dissipative fluid dynamics to electromagnetism. \ The former poses a formal challenge 
because, while it is well-known that non-dissipative fluid dynamics can be derived from an 
action principle \cite{taub54:_gr_variat_princ,carter89:_covar_theor_conduc,comer93:_hamil_multi_con,comer94:_hamil_sf,andersson07:_livrev}, the inclusion of dissipation in these systems 
tends to be phenomenological. \ The second is key if we want to move towards a greater 
level of realism in our astrophysics modeling. 

Given the first of these two issues, the recently proposed strategy for extending the 
variational approach to dissipative systems \cite{Andersson15:_dissfl_act} is promising. \ In 
principle, it provides us with an avenue for connecting dissipative channels with the 
underlying matter description and equation of state models accounting for transport 
phenomena. \ This paper aims to address the second issue by extending the variational 
derivation to account for electromagnetism. \ In particular, we provide a variational 
derivation for charged multifluid systems, accounting for  particle reactions and resistive 
scattering. \ Having obtained the formal results we discuss issues relating to 
electromagnetic gauge-invariance and develop a phenomenological model inspired by 
(and consistent with) the formal results. \ These developments provide a foundation for 
applications, as discussed in two companion papers 
\cite{Andersson16:_fibration,Andersson16:_foliation}. 

In Sec.~\ref{syskin}, we discuss the fundamental variables of the system and review the 
pull-back formalism. \ We also show how to build in general re-labeling invariance for the 
matter spaces. \ The total action, its variation, and resulting field equations are given in 
Sec.~\ref{actions}. \ In Sec.~\ref{pheres}, we use a decomposition of the system variables 
and fluid field equations with respect to a local ``observer's'' frame-of-reference to illuminate 
various features of the resistivity and to exploit them so as to produce a phenomenological 
model. \ A minimal model for  a system of electrons, protons, and heat is provided in 
Sec.~\ref{minmod}. \ Concluding thoughts and some discussion of immediate applications of 
the formalism are presented in Sec.~\ref{conclus}. \ Finally, in an appendix, we show how 
minimal coupling can be considered as a special type of entrainment between the 
electromagnetic four-potential and the charged fluid fluxes. \ The conventions of Misner, 
Thorne, and Wheeler \cite{mtw73} are used throughout.

\section{System Kinematics: The Fields and Variables}
\label{syskin}

We will assume that our system has a number $N_c$ of independent fluid constituents 
(such as electrons, protons, neutrons, and entropy). \ Each constituent has as its 
fundamental field a particle number density current $n^a_\x$, where $\x$ is a label that 
ranges over the various $N_c$ constituents ($\e$ for electrons, $\p$ for protons, etc.). \ The 
density $\nx$ associated with a given flux is given by $n^2_\x = - g_{a b} n^a_\x n^b_\x$. \ 
Among the $N_c$ constituents there will be a number $N_q$ which are charged, such that 
$N_q \leq N_c$. \ Each of these will have a charge $e_\x$ which combines with its 
associated flux current $n^a_\x$ to give a charged flux current  
$j^a_\x = e_\x n^a_\x$. \ Associated with each flux is a (canonically conjugate) fluid 
momentum $\mu^\x_a$ [\cf Eq.~\eqref{mux}]. \ While not dynamically independent (being a 
function of, in principle, all of the fluxes), its identification is an important component in 
extracting various physical properties of the system --- such as vorticity [\cf Eq.~\eqref{omx}]. 
\ The remaining field variables are the four-vector potential $A_a$ and the spacetime metric 
$g_{a b}$. \ With $A_a$ we couple the charged fluids to the electromagnetic field (and vice 
versa); the metric couples all fields to the spacetime curvature (and vice versa). \ At the end 
we have a complete system for describing a system of charged, gravitating, relativistic fluids.

\subsection{Fluid Particle Worldlines and Fluid Matter Space}

The (charged and uncharged) ``fluid particles'' associated with a given flux will have 
worldlines that follow from the unit four-velocity field $u^a_\x = n^a_\x/\nx$. \ In general, the 
number of independent four-velocities, or equivalently, the number of (charged and 
uncharged) fluids, $N_f$ will be equal to or less than $N_c$. \ This is determined from the 
outset by the details of the system that is to be described. \ When $N_f = N_c$, each 
constituent can move independently of the others, but when $N_f < N_c$, some of the 
constituents are flowing together; for example, as the limit of dynamical locking due to the resistive form of interaction 
developed later. 

\begin{figure}[t]
\centering
\includegraphics[height=9.2cm,clip]{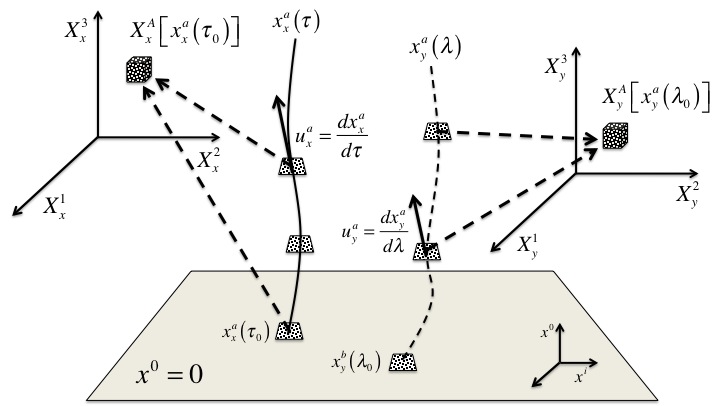}
\caption{A representation of the pull-back description for fluids based on matching 
worldlines in spacetime to points in matter space. \ We have placed on the worldlines small 
squares filled with dots. \ This is to emphasize the fact that the worldlines are for fluid 
elements, and not individual particles, and thus ``points'' on the worldlines are best thought 
of as small (with respect to the whole system) boxes containing a (thermodynamically 
describable) large number of particles.}
\label{2flpullbck}
\end{figure}

In Fig.~\ref{2flpullbck} we have a representation of some fluid-element worldlines, for a 
system of two fluids. \ With respect to the local coordinate system $\{x^0,x^i\}$, the points on 
the left-most, $\x$-fluid worldline are given by $x^a_\x\left(\tau\right)$, where $\tau$ is the 
proper time. \ The functions $x^a_\x\left(\tau\right)$ can be constructed from 
${\rm d} x^a_\x/{\rm d} \tau = u^a_\x$ once the fluid field equations are satisfied and 
$u^a_\x$ is known. \ Likewise, for the right-most, $\y$-fluid worldline, the functions 
$x^a_\y \left(\lambda\right)$, where $\lambda$ is the proper time, come from integrating 
${\rm d} x^a_\y/{\rm d} \lambda = u^a_\y$, once the $u^a_\y$ are known.  

With respect to system evolution, one often has in mind an initial-value approach to finding 
solutions. \ In our local coordinate system, we have an initial, spacelike slice at $x^0 = 0$, 
and so our determination of $x^a_\x\left(\tau\right)$ and $x^a_\y\left(\lambda\right)$ for the 
two worldlines will be based on the initial-value specification of their respective initial 
locations, namely, $x^a_\x\left(\tau_0\right)$ and $x^a_\y\left(\lambda_0\right)$. \ This brings 
up an important point: Assuming a given initial slice, and the fact that proper time allows for 
some re-parametrization invariance, we see that $x^i_\x\left(\tau_0\right)$ for each worldline 
is all that is needed to set up the distribution of the worldlines on the initial slice. \ In fact, it is 
plausible that once this labeling is in place, each fluid element will carry along with itself (via 
Lie-dragging) its original label as it moves along its worldline. 

This leads us to introduce for each fluid an abstract, three-dimensional ``matter'' space, the 
coordinates of which can be used as dynamical variables for the fluids. \ The role of the 
equations of motion is to guarantee that the initial set-up will lead to families of worldlines as 
fibrations of spacetime. \ On the left in Fig.~\ref{2flpullbck} we have placed the $\x$-fluid 
matter space having coordinates $X^A_\x$, $A = \{{\bf 1},{\bf 2},{\bf 3}\}$, and on the right is 
the $\y$-fluid matter space with coordinates $X^A_\y$. \ As we see in the figure, a ``point'' in 
either matter space is identified with a worldline of a fluid element in spacetime. \ The 
$X^A_\x$ are in general a set of three scalars on spacetime. \ They only vary from worldline 
to worldline, meaning for all $\tau$ of each worldline (and $\lambda$ for the other fluid 
worldlines)
\beq
      X^A_\x\left[\x^a_\x\left(\tau\right)\right] = X^A_\x\left[0,x^i_\x\left(\tau_0\right)\right] \ , 
\eeq
yet, for two different worldlines at, say, $\{0,x^i_\x\left(\tau_0\right)\}$ and 
$\{0,x^i_\x\left(\tau_0\right) + \delta x^i\}$, we have
\beq
      X^A_\x\left[0,x^i_\x\left(\tau_0\right)\right] 
                \neq 
      X^A_\x\left[0,x^i_\x\left(\tau_0\right) + \delta x^i\right] \ . 
\eeq
Next we will show how the $X^A_\x$ can be used as the fundamental fields for modeling 
fluid dynamics.

\subsection{Pull-back Formalism}

Consider the three-form $n^\x_{a b c}$ which is dual to $n^a_\x$:
\beq
    n^\x_{a b c} = \epsilon_{d a b c} n^d_\x 
                     \quad , \quad
    n^a_\x = \frac{1}{3!} \epsilon^{b c d a} n^\x_{b c d} \ , \label{n3form}
\eeq
where our convention for transforming between the two is
\beq
       \epsilon^{b c d a} \epsilon_{e b c d} = 3! \delta^a_e \ .
\eeq
Likewise, we introduce 
\beq
    \mu_\x^{a b c} = \epsilon^{d a b c} \mu^\x_d 
                     \quad , \quad
    \mu^\x_a = \frac{1}{3!} \epsilon_{b c d a} \mu_\x^{b c d} \ , \label{mu3form}
\eeq
which is the three-form dual to $\mu^\x_a$. 

We use the $X^A_\x$ to ``pullback'' $n^\x_{a b c}$ into the matter space where it is identified 
with $n^\x_{A B C}$: 
\beq
    n^\x_{a b c} = \jacx n^\x_{A B C} \ , \label{pb3form}
\eeq 
such that the Einstein convention applies to repeated matter space indices, and
\beq
      \jacx = \frac{\partial X^{[A}_\x}{\partial x^a} \frac{\partial X^B_\x}{\partial x^b} 
                   \frac{\partial X^{C]}_\x}{\partial x^c} \ .
\eeq
We also use the $X^A_\x$ to ``push-forward'' a matter space quantity, $\mu_\x^{A B C}$, to 
the spacetime three-form $\mu_\x^{a b c}$: 
\beq
       \mu_\x^{A B C} = \jacx \mu_\x^{a b c} \ .
\eeq
Note that this construction leads to $X^A_\x$ which are conserved along their own 
worldlines (\ie they are Lie-dragged by their $u^a_\x$):
\beq
        \frac{{\rm d} X^A_\x}{{\rm d} \tau_\x} = u^a_\x \nabla_a X^A_\x = \frac{1}{n_\x} \left(- 
        \frac{1}{3!} \epsilon^{a b c d} \frac{\partial X^A_\x}{\partial x^{[a}} 
        \frac{\partial X^B_\x}{\partial x^b} \frac{\partial X^C_\x}{\partial x^c} 
        \frac{\partial X^D_\x}{\partial x^{d]}}\right) n^\x_{B C D} \equiv 0 \ ,
\eeq
since the term in parentheses vanishes identically.

Because of the antisymmetry in its indices, $n^\x_{A B C}$ allows a natural definition 
for a volume-form $\epsilon^\x_{A B C}$ --- up to a normalization convention to be 
established in the next subsection --- on the $\x$-matter space:
\beq
       n^\x_{A B C} = \mathcal{N}^\x \epsilon^\x_{A B C} \ ,
\eeq
where $\mathcal{N}^\x$ will be defined momentarily. \ Similarly, the antisymmetry of the 
indices of $\mu_\x^{A B C}$ leads to an ``inverse'' volume form; namely, 
\beq
       \mu_\x^{A B C} = \mathcal{M}_\x \epsilon_\x^{A B C} \ ,
\eeq
where $\mathcal{M}_\x$ will also be defined momentarily. \ The quantity 
$\epsilon_\x^{A B C}$ is inverse in the sense that we impose
\beq
      \epsilon^\x_{D E F} \epsilon^{A B C}_\x = 3! \delta^{[A}_D \delta^B_E \delta^{C]}_F
              \ , \label{epsxforms}
\eeq
which implies
\beq
       \epsilon^\x_{A B C} \epsilon^{A B C}_\x = 3! \ . \label{epsabcsum}
\eeq
Now, we can write
\beq
       \mathcal{N}^\x = \frac{1}{3!} \epsilon^{A B C}_\x n^\x_{A B C} 
                              \quad , \quad
       \mathcal{M}_\x = \frac{1}{3!} \epsilon_{A B C}^\x \mu_\x^{A B C} \ .
\eeq
Letting 
\beq
       \nx = - n^a_\x u_a^\x 
                 \quad , \quad
       \mu^\x = - u^a_\x \mu^\x_a \ ,
\eeq
we have
\beq
       \mu^\x \nx = \mathcal{M}_\x \mathcal{N}^\x \ .
\eeq

\subsection{Matter Space Metrics}

To complete the establishment of $\epsilon^\x_{A B C}$ as a volume measure on its 
associated matter space, we need to establish normalizations for it and 
$\epsilon_\x^{A B C}$. \ Because of their antisymmetry, $\epsilon^\x_{A B C}$ and 
$\epsilon^{A B C}_\x$ each have only one independent component: $\epsilon^\x_{1 2 3}$ 
and $\epsilon^{1 2 3}_\x$, respectively. \ Establishing a normalization for 
$\epsilon^\x_{A B C}$ and $\epsilon^{A B C}_\x$ means setting values for 
$\epsilon^\x_{1 2 3}$ and $\epsilon^{1 2 3}_\x$. \ We will use a standard, linear algebra 
approach (see, Strang \cite{strang80:_lin_alg}) which, among other things, leads to 
re-labeling invariance for the matter spaces.

Note that the particle number densities can now take the form
\beq
       n^2_\x = \left(\sqrt{g_\x} \mathcal{N}^\x\right)^2 \ , 
\eeq
where
\beq
     g_\x = \frac{1}{3!} \epsilon^\x_{A B C} \epsilon^\x_{D E F} \hxad \hxbe \hxcf  
                 \quad , \quad \hxab 
              = \frac{\partial X^A_\x}{\partial x^a} \frac{\partial X^B_\x}{\partial x^b} g^{a b} \ ,
                  \label{hxab}
\eeq
and   
\beq
       n^2_{\x \y} = g_{\x \y} \mathcal{N}^\x \mathcal{N}^\y \ , 
\eeq
where
\beq
      g_{\x \y} = \frac{1}{3!} \epsilon^\x_{A B C} \epsilon^\y_{D E F} \hxyad \hxybe \hxycf 
                         \quad , \quad     
         \hxyab = \frac{\partial X^A_\x}{\partial x^a} \frac{\partial X^B_\y}{\partial x^b} g^{a b} \ . 
                         \label{hxyab}
\eeq
We will use the determinants of $\hxab$ and its inverse to form normalizations for 
$\epsilon^\x_{A B C}$ and $\epsilon^{A B C}_\x$.

The standard, matrix definition \cite{strang80:_lin_alg} for the determinant of $\hxab$ is
\beq
       \Delta_\x = \frac{1}{3! \left(\epsilon_{\rm 1 2 3}^\x\right)^2} \epsilon_{A B C}^\x
                           \epsilon_{D E F}^\x \hxad \hxbe \hxcf \ .
\eeq
The ``matrix'' inverse $\hxinvab$ of $\hxab$ is the solution to
\beq
      \hxac \hxinvcb = \hxca \hxinvcb = \hxinvbc \hxca = \hxinvbc \hxac = \delta^A_B \ , 
      \label{hxinv}
\eeq
and its determinant is 
\beq
       \Delta^\x = \frac{1}{3! \left(\epsilon^{\rm 1 2 3}_\x\right)^2} \epsilon^{A B C}_\x
                           \epsilon^{D E F}_\x \hxinvad \hxinvbe \hxinvcf \ .
\eeq
Our last step is to impose
\bea
      \epsilon^{A B C}_\x \epsilon^{D E F}_\x \hxinvad \hxinvbe \hxinvcf = 
      \epsilon_{A B C}^\x \epsilon_{D E F}^\x \hxad \hxbe \hxcf = 3! \label{norm} 
\eea
(which means $g_\x = 1$) and thus find
\beq
       \epsilon_{\rm 1 2 3}^\x = \frac{1}{\epsilon^{\rm 1 2 3}_\x} = \sqrt{\Delta^\x} = 
                       \frac{1}{\sqrt{\Delta_\x}}  \ .
\eeq

We will assume that the explicit form for $\hxinvab$ must be a combination of $\hxab$ 
and $\epsilon^\x_{A B C}$. \ Because $\hxab$ is symmetric, and taking into account 
Eq.~\eqref{epsxforms}, the only combination is
\beq
        \hxinvab = a_\x \epsilon^\x_{A C E} \epsilon^\x_{B D F} \hxcd \hxef \ .
\eeq
To complete the solution, we note that
\beq
        \hxab \hxinvab = \delta^A_A 
                                        \Longrightarrow
        a_\x = \frac{1}{2} \ .
\eeq
It is straightforward to verify that
\beq
       \hxac \left(\frac{1}{2} \epsilon^\x_{C D F} \epsilon^\x_{B E G} \hxde \hxfg\right) =  0 
                                   \quad , \quad
      A \neq B \ . 
\eeq

In a similar manner, we can find the inverse for $\hxyab$. \ This is a bit trickier, as we 
are mixing coordinates of two different matter spaces. \ We will consider it to have a 
``left''- and a ``right''-inverse:
\bea
     \hxyinvbc \hyxca &=& \hxyinvbc \hxyac = \delta^A_B \ , \cr
     \hxyac \hyxinvcb &=& \hyxca \hyxinvcb = \delta^A_B \ . \label{hxyinv}
\eea
As before, we use $\epsilon^{A B C}_\x$ and $\epsilon^\x_{A B C}$ to calculate the 
determinants
\bea
  \Delta_{\x \y} &=& \frac{1}{3! \epsilon_{\rm 1 2 3}^\x \epsilon_{\rm 1 2 3}^\y} 
                                  \epsilon_{A B C}^\x \epsilon_{D E F}^\y \hxyad \hxybe \hxycf 
                              = \Delta_{\y \x} \ , \cr
  \Delta^{\x \y} &=& \frac{1}{3! \epsilon^{\rm 1 2 3}_\x \epsilon^{\rm 1 2 3}_\y} 
                                  \epsilon^{A B C}_\x \epsilon^{D E F}_\y \hxyinvad \hxyinvbe \hxyinvcf 
                               = \Delta^{\y \x} \ .
\eea
The solution for the left-inverse is
\beq
        \hxyinvab = a^l_{\x \y} \epsilon^\y_{A C E} \epsilon^\x_{B D F} \hyxcd \hyxef \ ,
\eeq
where
\beq
        \hxyinvab \hyxba  = \delta^A_A 
                                        \Longrightarrow
        a^l_{\x \y} = \frac{1}{2} \frac{\Delta_{\y \x}}{\sqrt{\Delta_\y \Delta_\x}} \ .
\eeq
For the right-inverse it is
\beq
        \hyxinvab = a^r_{\y \x} \epsilon^\y_{A C E} \epsilon^\x_{B D F} \hyxcd \hyxef \ ,
\eeq
where
\beq
        \hxyab \hyxinvba   = \delta^A_A 
                                        \Longrightarrow
        a^r_{\x \y} = \frac{1}{2} \frac{\Delta_{\x \y}}{\sqrt{\Delta_\x \Delta_\y}} \ .
\eeq
Finally, we see that
\beq
       \hxyinvab = \hyxinvba \ .
\eeq

\subsection{Matter Space Covariance}

Because of the way we set up the worldline labeling --- they are assigned, arbitrarily, on 
some timelike slice (\cf Fig.~\ref{2flpullbck}) --- we can assert that there should be a 
relabeling invariance in the pull-back formalism. \ To that end, suppose we choose a new 
labeling scheme; \eg we use three scalars $Y^A_\x$ to mark individual fluid worldlines. \ 
However, this process must be constrained in the sense that it only changes the label of a 
given worldline, and does not map to a different worldline. \ Clearly this process is a 
mapping $F^A_\x$ from the $X^A_\x$ to the $Y^A_\x$; \ie
\beq
       Y^A_\x = F^A_\x\left\{X^B_\x\right\} \ .
\eeq 
Thus, the re-labeling of a worldline can be done, say, at $\{0,x^i_\x\left(\tau_0\right)\}$, 
where
\beq
       Y^A_\x\left[0,x^i_\x\left(\tau_0\right)\right] = 
       F^A\left\{X^B_\x\left[0,x^i_\x\left(\tau_0\right)\right]\right\} \ .
\eeq
Finally, the constancy of the $Y^A_\x$ along the worldline is preserved by the mapping 
since
\beq
        \frac{{\rm d} Y^A_\x}{{\rm d} \tau} = \frac{\partial F^A_\x}{\partial X^B_\x} 
                            \frac{{\rm d} X^B_\x}{{\rm d} \tau} = 0 \ .
\eeq

In principle, the $n_{A B C}^\x$ can have a functional dependence, say, of the $X^A_\x$ 
for each of the $N_c$ constituents as well as all the $\hxab$ and $\hxyab$. \  The
mapping $F^A_\x$ for the worldline at $\{0,x^i_\x\left(\tau_0\right)\}$ must be such that
\beq
       n^\x_{a b c}\left\{Y^A_\x\left[0,x^i_\x\left(\tau_0\right)\right],...\right\} = 
       n^\x_{a b c}\left\{X^A_\x\left[0,x^i_\x\left(\tau_0\right)\right],...\right\} \ , \label{nxmap}
\eeq
where the new matter space metric components are
\bea
       \bar{g}^{A B}_\x &\equiv& \frac{\partial Y^A_\x}{\partial x^a} 
                           \frac{\partial Y^B_\x}{\partial x^b} g^{a b} \cr
                  &=& \frac{\partial F^A_\x}{\partial X^C_\x} \frac{\partial F^B_\x}{\partial X^D_\x} 
                           \frac{\partial X^C_\x}{\partial x^a} \frac{\partial X^D_\x}{\partial x^b} 
                           g^{a b} \cr
                  &=& \frac{\partial F^A_\x}{\partial X^C_\x} \frac{\partial F^B_\x}{\partial X^D_\x} 
                           \hxcd 
\eea
and
\beq
        \bar{g}^{A B}_{\x \y} \equiv \frac{\partial Y^A_\x}{\partial x^a} 
                \frac{\partial Y^B_\y}{\partial x^b} g^{a b} 
             = \frac{\partial F^A_\x}{\partial X^C_\x} \frac{\partial F^B_\y}{\partial X^D_\y} \hxycd \ . 
\eeq
By rewriting Eq.~\eqref{nxmap}, we find
\bea
       0 &=& \bar{n}_{A B C}^\x \frac{\partial Y^A_\x}{\partial x^a} 
                   \frac{\partial Y^B_\x}{\partial x^b} \frac{\partial Y^C_\x}{\partial x^c} -
                   n_{A B C}^\x \frac{\partial X^A_\x}{\partial x^a} \frac{\partial X^B_\x}{\partial x^b} 
                    \frac{\partial Y^C_\x}{\partial x^c_\x} \cr
          &=& \left(\frac{\partial F^D_\x}{\partial X^A_\x} \frac{\partial F^E_\x}{\partial X^B_\x} 
                   \frac{\partial F^F_\x}{\partial X^C_\x} \bar{n}_{D E F}^\x - n_{A B C}^\x\right)
                   \frac{\partial X^A_\x}{\partial x^a} \frac{\partial X^B_\x}{\partial x^b_\x} 
                   \frac{\partial X^C_\x}{\partial x^c} \cr
          &\Longrightarrow& n_{A B C}^\x = \frac{\partial F^D_\x}{\partial X^A_\x} 
                   \frac{\partial F^E_\x}{\partial X^B_\x} \frac{\partial F^F_\x}{\partial X^C_\x} 
                   \bar{n}_{D E F}^\x \ ,
\eea
where
\beq
      \bar{n}_{A B C}^\x = \bar{n}_{A B C}^\x\left(F^A_\x,\frac{\partial F^A_\x}{\partial X^C_\x}  
                                           \frac{\partial F^B_\x}{\partial X^D_\x}\hxcd,...\right) \ .
\eeq

It has been asserted that Galilean invariance does not allow for $X^A_\x$ (via 
$\mathcal{N}_\x$) dependence in $n^\x_{A B C}$. \ But, we see that general mappings exist 
which preserve the covariance of the description, even those of a ``translation'' in matter 
space. \ This is important for what follows later, since the resistivity enters precisely because 
we allow for the full set of $n^\x_{A B C}$ to depend, in principle, on all of the $X^A_\x$.

\section{The Action Principle, Field Equations, and Gauge Invariance}
\label{actions}

In this section we will set up an action principle to derive the resistive-fluid, Maxwell, and 
Einstein set of field equations. \ The pull-back formalism will be used to set up variations of 
the fluid fluxes $n^a_\x$ required to get the fluid equations with resistivity. \ The Maxwell 
equations are obtained by varying $A_a$, which appears in two pieces of the total action: 
one built from the antisymmetric, Faraday tensor $F_{a b}$, defined as 
\beq
    F_{a b} = \nabla_a A_b - \nabla_b A_a \ ,
\eeq
and which satisfies a ``Bianchi'' identity
\beq
    0 = \nabla_a F_{b c} + \nabla_c F_{a b} + \nabla_b F_{c a} \ , \label{fbianchi}
\eeq
and the other constructed from a coupling term based on the scalar $j^a_\x A_a$. \ Finally, 
the stress-energy tensor $T_{a b}$ is obtained in the usual way by varying the total action 
with respect to the metric $g_{a b}$.

\subsection{The Matter, Electromagnetic, and Coupling Actions}

The fluid action $S_M$ (ignore boundary terms throughout) has as its Lagrangian the 
so-called Master function $\Lambda$, which depends on the 
$n^2_\x = - n^\x_a n^a_\x$ and the $n^2_{\x \y} = - g_{a b} n^a_\x n^b_\y$  (see 
\cite{andersson07:_livrev}). \ An arbitrary variation of $S_M$ with respect to the fluxes 
$n^a_\x$ and the metric gives 
\bea
    \delta S_M &=& \delta \left(\fint \Lambda\right) \cr
    &=& \fint \left[\sum_\x \mu^\x_a \delta n^a_\x + \frac{1}{2} \left(\Lambda g^{a b} + 
    \sum_\x n^a_\x \mu^b_\x\right) \delta g_{a b}\right] \ , \label{varlamb}
\eea
where $g$ is the determinant of the metric and $\mu^\x_a$ is the canonically 
conjugate momentum to $n^a_\x$; that is, letting
\beq
   \Bx = - 2 \frac{\partial \Lambda}{\partial n^2_\x} 
       \quad , \quad 
   \Axy = - \frac{\partial \Lambda}{\partial n^2_{\x \y}} \ ,
\eeq
then
\beq 
   \mu^\x_a = g_{a b} \left(\Bx n^b_\x + \sum_{\y \neq \x} \Axy 
              n^b_\y\right) \ . \label{mux}
\eeq

As mentioned earlier, the momentum is an essential piece of the formalism. \ For example, 
the antisymmetric vorticity two-form $\omega^\x_{a b}$ is obtained as the exterior derivative 
of $\mu^\x_a$; that is, 
\beq
   \omega^\x_{a b} = 2 \nabla_{[a} \mu^\x_{b]} \ . \label{omx}
\eeq
Its role as vorticity is well established; \eg when $\mu^\x_a$ is the gradient of a scalar then 
$\omega^\x_{a b}$ is zero (as is the case for superfluids). \ Notice also how the inclusion of 
$n^2_{\x \y}$ has led to so-called ``entrainment'',  a tilting of the fluid momenta in the 
sense that $\mu^\x_a$ is no longer simply proportional to its own flux $n^a_\x$. \ 
Entrainment
\cite{andreev75:_three_velocity_hydro,borumand96:_superfl_neutr_star_matter,comer03:_rel_ent,chamel06:_ent_cold_ns} 
between neutrons and protons is generally thought to be important in superfluid neutron 
stars and entrainment between matter and entropy can be shown to be important for causal 
heat conductivity \cite{Andersson10:_causal_heat}.   

The Maxwell Action is 
\beq
      S_{Max} = \frac{1}{16 \pi} \fint F_{a b} F^{a b} \ , 
\eeq
and its variation with respect to $A_a$ and the metric $g_{a b}$ is
\beq
    \delta S_{Max} = \frac{1}{4 \pi} \fint \left(\nabla_a F^{a b}\right) \delta A_b - 
               \frac{1}{32 \pi} \fint \left(F_{c d} F^{c d} g^{a b} - 4 F^{a c} F^b{}_c\right) 
               \delta g_{a b} \ .
\eeq
The minimal coupling of the Maxwell field to the charge current densities is obtained from 
the Coulomb action
\beq
    S_C = \fint \sum_\x j^a_\x A_a \ , \label{coulact}
\eeq
whose variation with respect to $n^a_\x$, $A_a$, and $g_{a b}$ gives 
\beq
    \delta S_C = \fint \sum_\x \left(j^a_\x \delta A_a + e_\x A_a \delta n^a_\x + \frac{1}{2} 
                            j^a_\x A_a g^{b c} \delta g_{b c}\right) \ .
\eeq
The variation of the total action $S$ for the system is thus
\bea
    \delta S &=& \delta S_M + \delta S_{Max} + \delta S_C \cr
             &=& \fint \left\{\sum_\x \left(\mu^\x_a + e_\x A_a\right) \delta n^a_\x + \frac{1}{4 \pi} 
                      \left(\nabla_b F^{b a} + 4 \pi \sum_\x j^a_\x\right) \delta A_a \right. \cr   
             &+& \left. \frac{1}{2} \left[\Lambda g^{a b} + \sum_\x \left(n^a_\x \mu^b_\x + 
                      j^c_\x A_c g^{a b}\right) - \frac{1}{16 \pi} \left(F_{c d} F^{c d} g^{a b} - 4 F^{a c} 
                      F^b{}_c\right)\right] \delta g_{a b}\right\} \ . \label{fmcvar}
\eea
Note that the minimal coupling has given a modification of the conjugate momentum familiar from, say, quantum mechanics; namely,
\beq
    \tilde{\mu}^\x_a = \mu^\x_a + e_\x A_a \ . 
\eeq

Of course, the field equations obtained from the variation above cannot be the final form, 
since the term proportional to $\delta n^a_\x$ implies that the momentum $\tilde{\mu}^\x_a$ 
must vanish. \ This occurs because the components of $\delta n^a_\x$ cannot all be varied 
independently. \ However,  the pull-back formalism provides  
a set of alternative variables, the $X^A_\x$, which can be varied independently. \ However, we have to 
incorporate the fact that the fluid momentum has changed from $\mu^\x_a$ to 
$\tilde{\mu}^\x_a$. \ This is straightforward since all that is required is to take 
Eq.~\eqref{mu3form} and replace $\mu^\x_a$ with $\tilde{\mu}^\x_a$, $\mu^\x_{a b c}$ with 
$\tilde{\mu}^\x_{a b c}$, and use that, as well as $\tilde{\mu}^\x = - u^a_\x \tilde{\mu}^\x_a$, 
as the basis for what follows below.

\subsection{Lagrangian Displacements}

Even though we have as our unconstrained dynamical variables the scalars $X^A_\x$, 
ultimately we want the action principle to produce field equations for the fluxes, since there 
is decades of literature and computational techniques for fluids based on solving for the 
components $n^a_\x$ and not the $X^A_\x$. \ Fortunately, we can use so-called 
Lagrangian displacements to bridge variations of matter space variables to those of 
spacetime. \ Denoted $\xi^a_\x$, the Lagrangian displacement for a fluid needs to be such 
that it tracks the virtual displacements of fluid element worldlines in spacetime. 

Using the standard definition of a Lagrangian variation in the relativistic context \cite{andersson07:_livrev}, we 
write 
\beq
        \Delta_\x X_\x^A = \delta X_\x^A + \mathcal{L}_{\xi_\x} X_\x^A = 0 \ , 
                                          \label{DelX}
\eeq
where $\delta X^A_\x$ is the Eulerian variation and $\mathcal{L}_{\xi_\x}$ is the Lie 
derivative. \ This means that convective variations are such that
\beq
    \delta X^A_\x = -  \mathcal{L}_{\xi_\x} X_\x^A = - \xi^a_\x 
                                \frac{\partial X^A_\x}{\partial x^a} \ . \label{xlagfl}
\eeq 
The displacements of the matter space fluid elements will lead to variations of 
$\delta n^\x_{A B C}$, which, in turn, will induce variations of $n^\x_{a b c}$. \ The existence 
of more than one fluid means, also, that we need to consider
\beq
    \Delta_\x X_\y^A = \delta X_\y^A + \lefx X_\y^A = \lefx X_\y^A - \lefy X_\y^A = \left(\xi^a_\x - 
                                      \xi^a_\y\right) \frac{\partial X^A_\y}{\partial x^a} \ .
\eeq

The Lagrangian variation of $n^\x_{a b c}$ in general is
\beq
   \Delta_\x n^\x_{a b c} = \jacx \Delta_\x n^\x_{A B C} \ , \label{nxgenvar}
\eeq
and thus
\beq
   \delta n^\x_{a b c} = - \lefx n^\x_{a b c} + \jacx \Delta_\x n^\x_{A B C} \ , \label{nxvargen}
\eeq
where the Lie derivative of the $n^\x_{a b c}$ along the $\xi^a_\x$ is
\beq
       \lefx n^\x_{a b c} = \xi^d_\x \frac{\partial n^\x_{a b c}}{\partial x^d}  
                        + n^\x_{d b c} \frac{\partial \xi^d_\x}{\partial x^a} 
                        + n^\x_{a d c} \frac{\partial \xi^d_\x}{\partial x^b} 
                        + n^\x_{a b d} \frac{\partial \xi^d_\x}{\partial x^c} \ .
\eeq

Andersson and Comer \cite{Andersson15:_dissfl_act} have demonstrated that allowing 
$n^\x_{A B C}$ to be a function of all the $X^A_\x$ (meaning include $X^A_\y$ for 
$\y \neq \x$), all the $\hxab$, and all the $\hxyab$ leads to a system of fluid equations with 
dissipation of several types, among which is the resistive type of interactions to be explored 
here and others coming from shear and bulk viscosities. \ The resistive forms of dissipation 
are due to the presence of $X^A_\y$ (for $\y \neq \x$) in the $n^\x_{A B C}$, and so we 
consider here
\bea
     \Delta_\x n^\x_{A B C} &=& \sum_{\y \neq \x} \frac{\partial n^\x_{A B C}}{\partial X^D_\y} 
              \Delta_\x X^D_\y \cr
      &=& \sum_{\y \neq \x} \frac{\partial n^\x_{A B C}}{\partial X^D_\y} 
               \frac{\partial X^D_\y}{\partial x^a} \left(\xi^a_\x - \xi^a_\y\right) \ .  \label{Delnxabc}   
\eea

Using the facts that
\beq
   \Delta_\x g^{a b} = \delta g^{a b} - 2 \nabla^{(a} \xi^{b)}_\x \ , 
\eeq
\beq
   \delta \epsilon^{a b c d} = - \frac{1}{2} \epsilon^{a b c d} g^{e f} \delta g_{e f} \ , 
\eeq
and
\beq
      \epsilon^{b c d a} \lefx n^\x_{b c d} = 3! \left(\xi^b_\x \nabla_b n^a_\x - n^b_\x \nabla_b 
      \xi^a_\x + n^a_\x \nabla_b \xi^b_\x\right) \ ,
\eeq
we find
\bea
    \delta n^a_\x &=& \delta \left(\frac{1}{3!} \epsilon^{b c d a} n^\x_{b c d}\right) \cr
    &=& n^b_\x \nabla_b \xi^a_\x - \xi^b_\x \nabla_b n^a_\x - n^a_\x \left(\nabla_b \xi^b_\x + 
             \frac{1}{2} g^{b c} \delta g_{b c}\right) + \frac{1}{\tilde{\mu}^\x n_\x} n^a_\x 
             \sum_{\y \neq \x} \RXYb \left(\xi^b_\x - \xi^b_\y\right) \ ,
\eea
where
\beq
   \RXYa \equiv \frac{1}{3!} \frac{\partial X^D_\y}{\partial x^a} \tilde{\mu}_\x^{A B C} 
            \frac{\partial n^\x_{A B C}}{\partial X^D_\y} \label{RXYa} 
\eeq
and it satisfies the identity
\beq
        u^a_\y \RXYa \equiv 0 \ . \label{rxyrest}
\eeq
The total ``resistivity'' current $\rtotxa$ is
\beq
    \rtotxa = \sum_{\y \neq \x} \left(\RYXa - \RXYa\right) \ , \label{rxdef} 
\eeq
which has the identity
\beq
        \sum_\x \rtotxa \equiv 0 \ . \label{rtotvan}
\eeq

\subsection{The Field Equations}

We now return to the flux variations of the total action given in Eq.~\eqref{fmcvar}. \ The fact 
that we are summing over all constituents leads to  
\beq
      \sum_\x \sum_{\y \neq \x} \RXYa \left(\xi^a_\x - \xi^a_\y\right) 
      = \sum_\x \sum_{\y \neq \x} \left(\RXYa - \RYXa\right) \xi^a_\x 
      = - \sum_\x \rtotxa \xi^a_\x \ ,
\eeq
so that the variation of the total action for the system is 
\bea
    \delta S &=& \fint \left\{- \sum_\x \left(f^\x_a + \Gamma_\x \tilde{\mu}^\x_a - \rtotxa\right) 
                           \xi^a_\x - \frac{1}{4 \pi} \left(\nabla_b F^{a b} - 4 \pi \sum_\x j^a_\x\right) 
                           \delta A_a \right. \cr 
                     && \left. + \frac{1}{2} \left[\Psi g^{a b} + \sum_\x n^a_\x \mu^b_\x - 
                           \frac{1}{16 \pi} \left(F_{c d} F^{c d} g^{a b} - 4 F^{a c} F^b{}_c\right)\right] 
                           \delta g_{a b}\right\} \ . 
\eea
where 
\beq
      f^\x_a = n^b_\x \tilde{\omega}^\x_{b a} \equiv 2 n^b_\x \nabla_{[a} \tilde{\mu}^\x_{b]} \ , 
\eeq
\beq
       \Gx = \nabla_a n^a_\x \ ,
\eeq
and
\beq
       \Psi = \Lambda - \sum_\x \mu^\x_c n^c_\x \ .
\eeq

The Euler equation for each fluid is
\beq
    f^\x_a + \Gx \tilde{\mu}^\x_a = \rtotxa \ , \label{xfleqn}
\eeq
the Maxwell equation (including also Eq.~\eqref{fbianchi}) is
\beq
    \nabla_b F^{a b} = \nabla_b \left(\nabla^a A^b - \nabla^b A^a\right) = 4 \pi \sum_\x 
    j^a_\x \ ,
\eeq
and the stress-energy tensor is
\beq
    T^{a b} = \Psi g^{a b} + \sum_\x n^a_\x \mu^b_\x - \frac{1}{16 \pi} \left(F_{c d} F^{c d} 
                     g^{a b} - 4 F^{a c} F^b{}_c\right) \ . \label{emstens}
\eeq

\subsection{Impact of Change of Gauge for $A_a$}

A gauge transformation will impact the fluid equations of motion because of the change to 
the momentum; \ie letting $\bar{A}_a = A_a + \nabla_a \phi$ we find
\beq
       \tilde{\mu}^\x_a = \mu^\x_a + e_\x A_a 
                                       \quad \longrightarrow \quad
       \bar{\mu}^\x_a = \mu^\x_a + e_\x \bar{A}_a = \tilde{\mu}^\x_a + e_\x \nabla_a \phi \ .
\eeq
It is important here to consider in more detail the ramifications of a change of gauge, since 
an application of the present work will be to numerical evolutions 
\cite{Andersson16:_foliation}. \ In the numerical setting, we expect to be solving for the 
vector potential $A_a$ as we evolve the system. \ This will require a choice of gauge for the 
vector potential, which will affect the explicit values of terms (such as the resistivity) in the 
equations of motion. 

In Eq.~\eqref{xfleqn} (\ie the fluid equation of motion), we see the term involving $\Gx$ is 
changed but not $f^\x_a$. \ What also changes is $\rtotxa$, since the quantity 
$\tilde{\mu}^{A B C}_\x$ in $\RXYa$ [\cf Eq.~\eqref{RXYa}] depends on $A_a$. \ Letting 
$\rtotxbara$ denote the resistivity in the new gauge, we find
\bea
 \rtotxbara &=& \sum_{\y \neq \x} \left(\RYXbara - \RXYbara\right) \cr
                   &=& \sum_{\y \neq \x} \frac{1}{3!} \epsilon^{e b c d} \left[\left(\tilde{\mu}^\y_e + 
                           e_\y \nabla_e \phi\right) \jacybcd \frac{\partial n^\y_{A B C}}{\partial X^D_\x} 
                           \frac{\partial X^D_\x}{\partial x^a} \right. \cr
                     && \left. - \left(\tilde{\mu}^\x_e + e_\x \nabla_e \phi
                           \right) \jacxbcd \frac{\partial n^\x_{A B C}}{\partial X^D_\y} 
                           \frac{\partial X^D_\y}{\partial x^a}\right] \cr
                  &=& \rtotxa + \gtotxa \ ,  
\eea
where
\beq
       \gtotxa = \sum_{\y \neq \x} \left(\GYXa - \GXYa\right) 
                        \quad , \quad
       \GXYa = \frac{1}{3!} \epsilon^{e b c d} e_\x \left(\jacxbcd 
                        \frac{\partial n^\x_{A B C}}{\partial X^D_\y} \frac{\partial X^D_\y}{\partial x^a}
                        \right) \nabla_e \phi \ . \label{gxya}
\eeq
Note that 
\beq
     \sum_\x \rtotxa = \sum_\x \gtotxa = 0 
                           \quad \Longrightarrow \quad 
     \sum_\x \rtotxbara = \sum_\x \rtotxa + \sum_\x \gtotxa = 0 \ .
\eeq
In the new gauge the fluid equation of motion becomes
\bea
        0 &=& \bar{f}^\x_a + \Gx \bar{\mu}^\x_a + \rtotxbara \cr
           &=& f^\x_a + \Gx \left(\tilde{\mu}^\x_a + e_\x \nabla_a \phi\right) - 
                    \left(\rtotxa + \gtotxa\right) \ .
\eea
Projecting along $n^a_\x$ we find
\bea
        0 &=& \Gx n^a_\x \left(\tilde{\mu}^\x_a + e_\x \nabla_a \phi\right) - n^a_\x \left(\rtotxa + 
                    \gtotxa\right) \ .
\eea

We have seen above that the observables, including the stress-energy tensor, Faraday 
tensor, and all hydrodynamic variables are independent of the choice of gauge for $A_a$, 
as expected. \ However, the fluid field equations are modified, which is also expected. \ 
Nevertheless, we can determine the modifications and thereby evolve the system 
regardless of the choice of gauge.

\subsection{Gauge Invariance and Charge Conservation}

To see other consequences of gauge invariance, we will consider a variation of the total 
action, where the vector potential variation takes the form
\beq
       \delta A_a = \nabla_a \delta \phi \ .
\eeq
We assume that $\xi^a_\x = 0$ and $\delta g_{a b} = 0$ under the change of gauge; thus, 
even though the term $\RXYa$ acquires the gauge term $\GXYa$ [\cf Eq.~\eqref{gxya}] it 
does not affect $\delta n^a_\x$. \ The total action thus reduces to
\bea    
  \delta S &=& - \frac{1}{4 \pi} \fint \left(\nabla_b F^{a b} - 4 \pi \sum_\x j^a_\x\right) 
                          \nabla_a \delta \phi  \cr 
                &=& - \frac{1}{4 \pi} \fint \nabla_a \left(\nabla_b F^{a b} - 4 \pi \sum_\x 
                         j^a_\x\right) \delta \phi \ , 
\eea
which implies
\beq
      \nabla_a \left(\nabla_b F^{a b}\right) = 4 \pi \sum_\x e^\x \nabla_a n^a_\x 
      = 4 \pi \sum_\x e^\x \Gx \ .
\eeq
However, the commutation of covariant derivatives acting on a two-index object is
\beq
         \nabla_a \nabla_b F^c{}_d - \nabla_b \nabla_a F^c{}_d = R^c{}_{e a b} F^e{}_d - 
                   R^e{}_{d a b} F^c{}_e \ ;
\eeq
hence,
\beq
  \frac{1}{4 \pi} \nabla_a \left(\nabla_b F^{a b}\right) = \frac{1}{4 \pi} R_{a b} F^{a b} \equiv 0 \ , 
\eeq
since the Ricci tensor is symmetric and the Faraday tensor is antisymmetric. \ Thus, we 
recover the expected conservation of charge:
\beq
        \sum_\x e^\x \Gx = \sum_\x \nabla_a j^a_\x = 0 \ . \label{chcons}
\eeq 
Using the field equations, and Eqs.~\eqref{rtotvan} and \eqref{chcons}, we can show that 
$\nabla_a T^{a b}$ vanishes identically (as it should from diffeomorphism invariance):
\bea
      \nabla_b T^b{}_a &=& \nabla_b \left[\Psi \delta^b_a + \sum_\x n^b_\x \mu^\x_a - 
      \frac{1}{16 \pi} \left(F_{c d} F^{c d} \delta^b_a - 4 F^{b c} F_{a c}\right)\right] \cr
      &=& \sum_\x \rtotxa + \left(\sum_\x e_\x \Gx\right) A_a \equiv 0 \ .
\eea

\section{A phenomenological approach to the resistivity}
\label{pheres}

Having completed the formal considerations, we can turn our attention to applications. As we do so, 
it is very important to appreciate that the $n^\x_{A B C}$ and how they enter $\Lambda$ 
is understood to be ``known'' a priori. \ It is not until a specific application is intended that 
one would necessarily require an explicit relation. \ An analogy is the Lagrangian for an 
interacting complex scalar field. \ A potential $V(\phi^\dagger \phi)$ is introduced, but 
not generally specified until the Euler-Lagrange equations are derived and a specific 
application is pursued. 

At this point, the action principle has given us the tensorial structure 
of the equations and how many different dissipative processes exist in general. \ Ideally, 
what we would do next is use microphysics to specify the $n^\x_{A B C}$ and $\Lambda$. \ 
Admittedly that task is daunting and would require more specifics about the actual systems 
to be described. \ Instead, we will develop here a phenomenological form of the resistivity
$\rtotxa$, which is consistent with the field equations above, the various identities, and the 
Second Law of Thermodynamics.

To begin, it is convenient to introduce a fiducial frame-of-reference\footnote{In an 
accompanying paper \cite{Andersson16:_fibration}, we will consider a family of worldlines of 
this type and form a fibration of spacetime, and in \cite{Andersson16:_foliation} we will 
make use of a field $N^a$ which is surface-forming and hence can provide a foliation 
for a $3 + 1$ decomposition of spacetime.} whose worldline is determined by the unit 
four-velocity $u^a$. \ Locally, we can decompose our fields into pieces parallel to $u^a$ 
and perpendicular to $u^a$ using the projection operator
\beq
    \perp^a_b \equiv \delta^a_b + u^a u_b
                  \quad , \quad
    u_a u^a = - 1 \ .
\eeq
For instance, the particle flux unit vectors are now decomposed as
\beq
    v^a_\x \equiv \gx \perp^a_b u^b_\x
                  \quad \Longrightarrow \quad
    u^a_\x = \gx \left(u^a + v^a_\x\right) 
                    \quad , \quad 
    \gamma^2_\x = \left(1 - v^a_\x v^\x_a\right)^{- 1} \ , \label{uxdec}
\eeq
where $v^a_\x$ is the (coordinate-based time) three-velocity.

Recall that the resistivity is given by [\cf Eqs.~\eqref{rxyrest} and \eqref{rxdef}]
\beq
         R^\x_a = \sum_{\y \neq \x} \left(R^{\y \x}_a - R^{\x \y}_a\right) 
                          \quad , \quad
         u_\y^a R^{\x \y}_a = 0 \ , 
\eeq
and $\RXYa$ is defined in Eq.~\eqref{RXYa}. \ Its decomposition is 
\beq
      \hat{R}^{\x \y}_a \equiv \perp^b_a \RXYb
               \quad \Longrightarrow \quad
      \RXYa = \left(- u^b \RXYb\right) u_a + \hat{R}^{\x \y}_a \ .
\eeq
The constraint on $\RXYa$ [\cf Eq.~\eqref{rxyrest}] becomes
\bea
       0 &=& u^a_\y \RXYa = \gamma_\y \left(u^a + v^a_\y\right) \RXYa \cr
             && \Longrightarrow - u^a \RXYa =  v^a_\y \hat{R}^{\x \y}_a \ ;
\eea
thus, $\RXYa$ --- for given $\x$ and $\y$ --- has only $3$ free components 
$\hat{R}^{\x \y}_a$, and takes the form
\beq
        \RXYa = \left(\delta^b_a + v^b_\y  u_a\right) \hat{R}^{\x \y}_b \ .
\eeq
Putting all these pieces together, the total resistivity takes the form
\beq
        \rtotxa = \sum_{\y \neq \x} \left[\left(v^b_\x \hat{R}^{\y \x}_b - v^b_\y \hat{R}^{\x \y}_b
                       \right) u_a + \hat{R}^{\y \x}_a - \hat{R}^{\x \y}_a\right] \ . \label{rxdec}     
\eeq
It is easy to see that the ``time'' and ``space'' pieces separately satisfy Eq.~\eqref{rtotvan}.  

Using the fluid equations of motion we can relate the resistivity to the particle number 
creation rate $\Gx$. \ Note that Eq.~\eqref{rxdec} implies
\beq
      u_\x^a \rtotxa = \gx \left(u^a + v^a_\x\right) \rtotxa 
                               = - \gx \sum_{\y \neq \x} w^a_{\x \y} \hat{R}^{\x \y}_a \ ,
\eeq
where
\beq
      w^a_{\x \y} = v^a_\x - v^a_\y \ .
\eeq
A projection of the fluid field equation [\cf Eq.~\eqref{xfleqn}] along the $u_\x^a$ flow leads 
to
\beq
     \Gx  u_\x^a \tilde{\mu}^\x_a \equiv - \tilde{\mu}^\x \Gx =  
              u_\x^a \rtotxa \ , \label{fricon}
\eeq
so that
\beq
       \Gx = \left(\gamma^{- 1}_\x \tilde{\mu}^\x\right)^{- 1} \sum_{\y \neq \x} 
                                w^a_{\x \y} \hat{R}^{\x \y}_a \ . \label{gamcon}
\eeq
To further constrain the resistivity, we can use conservation of charge 
[\cf Eq.~\eqref{chcons}], overall charge neutrality, and the Second Law of Thermodynamics 
($\Gamma_\s \geq 0$). \ The conservation of charge implies
\beq
       0 = \sum_\x e^\x \Gx 
           = \sum_\x \frac{e^\x}{\gamma^{- 1}_\x \tilde{\mu}^\x} \sum_{\y \neq \x} 
                   w^a_{\x \y} \hat{R}^{\x \y}_a \ , \label{chcons2}
\eeq 
and the Second Law of Thermodynamics takes the form
\beq
       \Gamma_\s = \left(\gamma^{- 1}_\s \tilde{\mu}^\s\right)^{- 1} \sum_{\{\x \neq \s\}} 
                                w^a_{\s \x} \hat{R}^{\s \x}_a \geq 0 \ . \label{2ndlaw}
\eeq

We have not yet made any approximation in our system. \ However, our goal here is to 
produce a phenomenological model, and so it makes sense to now employ the standard 
analysis due to Onsager \cite{onsager31:_symmetry} (see also  
\cite{andersson06:_flux_con,bryn}). \ The point is to introduce a form for the dissipation by 
identifying thermodynamic fluxes --- here the $\hat{R}^{\x \y}_a$ --- and forces --- the 
$w^a_{\x \y}$. \ These quantities must be such that they tend to drive the system to 
equilibrium --- the fluids become comoving ($w^a_{\x \y} = 0$ for all $\x$ and $\y$) --- while 
simultaneously maintaining the inequality of Eq.~\eqref{2ndlaw}. 

Clearly, a model which makes the entropy production manifestly positive-definite will work 
and so we assume
\beq
         \hat{R}^{\x \y}_a = \bar{R}^{\x \y} w^{\x \y}_a 
                  \quad \Longrightarrow \quad
         \RXYa = \bar{R}^{\x \y} \left(\delta^b_a +  u_a v^b_\y\right) w^{\x \y}_b \ ,
\eeq
which leads to
\beq
        \rtotxa = \sum_{\y \neq \x} \left[\left(\bar{R}^{\y \x} v^b_\x + \bar{R}^{\x \y} v^b_\y\right) 
                        u_a + \left(\bar{R}^{\y \x} + \bar{R}^{\x \y}\right) \delta^b_a\right] w^{\y \x}_b \ .      
\eeq
Now introduce
\beq
       \mathcal{R}^{\x \y} = \bar{R}^{\y \x} + \bar{R}^{\x \y}
\eeq
(obviously symmetric in $\x$ and $\y$) to get 
\beq
        \rtotxa = \left(\sum_{\y \neq \x} \bar{R}^{\x \y} w^b_{\x \y} w^{\x \y}_b\right) u_a + 
                        \sum_{\y \neq \x} \mathcal{R}^{\x \y} \left(\delta^b_a + v^b_\x u_a\right) 
                        w^{\y \x}_b \ . 
\eeq
Noting that
\beq
        \Gx = \left(\gamma^{- 1}_\x \tilde{\mu}^\x\right)^{- 1} \sum_{\y \neq \x} 
                                 \bar{R}^{\x \y} w^b_{\x \y} w^{\x \y}_b \ ,
\eeq
we finally arrive at
\beq
         \rtotxa = \left(\gamma^{- 1}_\x \tilde{\mu}^\x \Gx\right) u_a + \sum_{\y \neq \x} 
                        \mathcal{R}^{\x \y} \left(\delta^b_a + v^b_\x u_a\right) w^{\y \x}_b \ .      
\eeq

If there are no reactions ($\Gx = 0$) then
\beq
       R^\x_a = \sum_{\y \neq \x} \mathcal{R}^{\x\y} (\delta_a^b + v_\x^b u_a) w^{\y \x}_b \ .
\eeq
Given that the resistivities can depend, in principle, on {\em all} of the fluids in the system, 
any restriction like zero particle creation for a subset of the fluids will have an impact on all 
the particle creation rates; in particular, the entropy creation rate.

\section{What is the Minimal Model that Includes Resistivity?}
\label{minmod}

Even with this more specific model, there are still a number of degrees of freedom --- the 
undetermined coefficients $\bar{R}^{\x \y}$; namely, if we have $N_c$ constituents, then for 
each choice of $\x$, there will be $N_c - 1$ choices for $\y$, and thus a maximum of 
$N_c \left(N_c - 1\right)$ coefficients. \ Note that the condition expressed in 
Eq.~\eqref{rtotvan} is satisfied identically and so it does not reduce the number of free 
$\bar{R}^{\x \y}$. \ The conservation of charge is another matter. \ Ideally, it is also an 
identity, meaning that the total action $S_{FMC}$ must be constructed in such a way that it 
incorporates the electromagnetic gauge symmetry. \ However, in our phenomenological 
model, we have chosen a form for the $\RXYa$ --- it has not been derived as in 
Eq.~\eqref{rxdef} --- and so we must impose charge conservation ``by hand'', meaning that 
Eq.~\eqref{chcons} is in fact an additional constraint on the system. \ Hence, a complete 
specification of the model will require knowing $N_c \left(N_c - 1\right) - 1$ of the 
$\bar{R}^{\x \y}$ coefficients.

We will first consider the simplest problem of a two-fluid, two-constituent system where the 
two types of particles have equal but opposite charges ($- e_\e = e_\p \equiv e$). \ The 
particle creation rates are
\bea
      \gamma^{- 1}_\e \tilde{\mu}^\e \Gamma_\e &=& \bar{R}^{\e \p} w_{\e \p}^2 \ , \\
      \gamma^{- 1}_\p \tilde{\mu}^\p \Gamma_\p &=& \bar{R}^{\p \e} w_{\p \e}^2 \ .      
      \label{2flplas}
\eea
(noting that $w_{\e\p}^a$ is spatial).
Note that charge conservation [\cf Eq.~\eqref{chcons2}] implies $\Gamma_\e = \Gamma_\p$, 
or
\beq
         \gamma^{- 1}_\p \tilde{\mu}^\p \bar{R}^{\e \p} = \gamma^{- 1}_\e \tilde{\mu}^\e 
                       \bar{R}^{\p \e} \ . \label{chrgcons}
\eeq
As the sum of $\rtotea$ and $\rtotpa$ vanishes identically, we see, as expected, that there is 
only one free component $\bar{R}^{\e \p}$. \ Finally, the two resistivities are
\bea
        \rtotea &=& \bar{R}^{\e \p} \left[w^2_{\e \p}u_a - \left(1 + 
                             \frac{\gamma^{- 1}_\p \tilde{\mu}^\p}{\gamma^{- 1}_\e \tilde{\mu}^\e}\right) 
                             w^{\e \p}_b \left(\delta^b_a + v^b_\e u_a\right)\right] \ , \\
        \rtotpa &=& \frac{\gamma^{- 1}_\p \tilde{\mu}^\p}{\gamma^{- 1}_\e \tilde{\mu}^\e}  
                             \bar{R}^{\e \p} \left[w^2_{\e \p} u_a + \left(1 + 
                             \frac{\gamma^{- 1}_\e \tilde{\mu}^\e}{\gamma^{- 1}_\p \tilde{\mu}^\p}\right) 
                             w^{\e \p}_b \left(\delta^b_a + v^b_\p u_a\right)\right] \ .      
\eea
However, many applications in plasma physics have zero particle creation rates, and we 
see in this case that the resistivities vanish. \ Essentially, we are proving that there can be no 
resistivity without also taking into account heat; \ie a non-zero entropy creation rate.

The simplest, non-trivial system has the two charged fluids and entropy. \ The creation rates 
expand to
\bea
      \gamma^{- 1}_\e \tilde{\mu}^\e \Gamma_\e &=& \bar{R}^{\e \p}  w^2_{\e \p} + 
                                    \bar{R}^{\e \s}  w^2_{\e \s} \ , \\
      \gamma^{- 1}_\p \tilde{\mu}^\p \Gamma_\p &=& \bar{R}^{\p \e} w^2_{\p \e} + 
                                    \bar{R}^{\p \s}  w^2_{\p \s} \ , \\
      \gamma^{- 1}_\s \tilde{\mu}^\s \Gamma_\s &=& \bar{R}^{\s \e}  w^2_{\s \e} +  
                                    \bar{R}^{\s \p}  w^2_{\s \p} \ , \label{3flplas}
\eea
and the resistivities are
\bea
        \rtotea &=& \left(\bar{R}^{\e \p} w^2_{\p \e} + \bar{R}^{\e \s} w^2_{\s \e} 
                             \right) u_a  + \left(\mathcal{R}^{\p \e} w^{\p \e}_b + 
                             \mathcal{R}^{\s \e} w^{\s \e}_b\right) \left(\delta^b_a + v^b_\e u_a\right) \ , \\
        \rtotpa &=& \left(\bar{R}^{\p \e} w^2_{\e \p}  + \bar{R}^{\p \s} w^2_{\s \p} 
                             \right) u_a + \left(\mathcal{R}^{\e \p} w^{\e \p}_b + 
                               \mathcal{R}^{\s \p} w^{\s \p}_b\right) \left(\delta^b_a + v^b_\p u_a\right) \ , \\
        \rtotsa &=& \left(\bar{R}^{\s \e} w^2_{\e \s}  + \bar{R}^{\s \p} w^2_{\p \s} 
                             \right) u_a + \left(\mathcal{R}^{\e \s} w^{\e \s}_b + 
                             \mathcal{R}^{\p \s} w^{\p \s}_b\right) \left(\delta^b_a + v^b_\s u_a\right) \ .      
\eea
Charge conservation gives
\bea
      0 &=& \left(\frac{1}{\gamma^{- 1}_\p \tilde{\mu}^\p} \bar{R}^{\p \e} - 
                  \frac{1}{\gamma^{- 1}_\e \tilde{\mu}^\e} \bar{R}^{\e \p}\right)  w^2_{\p \e} 
                  + \left(\frac{1}{\gamma^{- 1}_\p \tilde{\mu}^\p} \bar{R}^{\p \s} w^2_{\p \s} 
                  - \frac{1}{\gamma^{- 1}_\e \tilde{\mu}^\e} \bar{R}^{\e \s}  w^2_{\e \s}
                  \right) 
\eea 
and the Second Law [\cf Eq.~\eqref{2ndlaw}] implies
\beq
        \bar{R}^{\s \e} \ , \ \bar{R}^{\s \p} \geq 0 \ .
\eeq
If we now assume that there is no charge creation, then
\bea
     0 &=& \bar{R}^{\e \p}  w^2_{\e \p} + \bar{R}^{\e \s}  w^2_{\e \s} \ , \\
     0 &=& \bar{R}^{\p \e}  w^2_{\e \p} + \bar{R}^{\p \s}  w^2_{\p \s} \ . 
\eea
Unlike before, we can satisfy these conditions with something as simple as requiring the 
coefficients $\bar{R}^{\e \p}$, $\bar{R}^{\p \e}$, $\bar{R}^{\e \s}$, and $\bar{R}^{\p \s}$ to 
vanish.\footnote{Since $w^2_{\e \p}$, $w^2_{\e \s}$, and 
$w^2_{\p \s}$ are linearly independent, this is tantamount to assuming that the 
$\bar{R}^{\x \y}$ have negligible dependence on the relative velocities.} \ This will leave us 
with only two free coefficients, $\bar{R}^{\s \e}$ and $\bar{R}^{\s \p}$, and resistivities of the 
form
\bea
        \rtotea &=& \bar{R}^{\s \e} w^{\s \e}_b \left(\delta^b_a + v^b_\e u_a\right) \ , \\
        \rtotpa &=& \bar{R}^{\s \p} w^{\s \p}_b \left(\delta^b_a + v^b_\p u_a\right) \ , \\
        \rtotsa &=& \left(\bar{R}^{\s \e} w^2_{\e \s}  + \bar{R}^{\s \p} w^2_{\p \s} 
                             \right) u_a + \left(\bar{R}^{\s \e} w^{\e \s}_b + 
                             \bar{R}^{\s \p} w^{\p \s}_b\right) \left(\delta^b_a + v^b_\s u_a\right) \ .  
                             \label{phenmod}    
\eea

Perhaps the most important point of developing this kind of phenomenological model is to 
show that, even without specific forms for the $n^\x_{A B C}$ and $\Lambda$, the multi-fluid 
formalism is robust enough to build increasingly complex models without first having to 
perform microphysical calculations. \ Of course, we would still need some insight from 
microphysics; \eg to determine $\bar{R}^{\s \e}$ and $\bar{R}^{\s \p}$.

\section{Conclusions and Follow-on Work}
\label{conclus}

The relativistic fluid system is the backbone of modeling many systems in astrophysics, 
cosmology and high-energy physics. \ Here, we have taken a unique step in the 
development of the fluid modeling scheme: an action principle has been used to build a 
system of field equations for relativistic plasmas including resistivity. \ This is a ``first 
principles'' approach which is logically concise in the sense that many of the assumptions 
about the system's physics can be traced to the initial phase of constructing the action; in 
particular, it was straightforward to take the action principle for dissipative, relativistic fluids 
from \cite{Andersson15:_dissfl_act} and add to it the standard action for electromagnetic 
fields and the usual Coulomb coupling of the charged fluxes to the electromagnetic 
four-potential. 

The present discussion is complemented by two companion papers.
In \cite{Andersson16:_fibration} we use this work's results to develop a fully relativistic 
framework that allows for four (fluid) components: normal and superconducting currents, 
heat flow, and a final component with normal and superfluid flows. \ The purpose of the 
model is to make contact ---  in the appropriate limit --- with ideal magnetohydrodynamics. \ 
A key component of the framework is the insertion of a suitable family of observers of the 
fluid flow, who basically provide a fibration of spacetime.  \ While the model is suitable to 
describe isolated superfluid neutron stars, it is not appropriate for numerical simulations of 
(say) merging neutron stars. \ Progress in this direction is made in 
\cite{Andersson16:_foliation}, which connects with the present discussion through use of a 
3+1 foliation of spacetime. 

While our focus here was on the resistivity, there is a clear process for building on these 
results by adding in other dissipation channels (such as those arising from bulk and shear 
viscosities) already included in the action principle of Andersson and Comer
\cite{Andersson15:_dissfl_act}. \ Basically, we may follow the procedure presented here, with the 
only change being to include terms like the matter space metrics $\hxab$ and $\hxyab$ [\cf 
Eqs.~\eqref{hxab} and \eqref{hxyab}] in the variation of $n^\x_{AB C}$ [\cf 
Eq.~\eqref{Delnxabc}]. \ This speaks to the power of having a first principles approach to 
developing the overall form of the field equations, even if details of the formalism still will 
require microphysics for dissipation coefficients [such as $\bar{R}^{\s \e}$ and 
$\bar{R}^{\s \p}$ in the phenomenological model; \cf Eq.~\eqref{phenmod}].

To conclude, the variational approach has allowed us to make significant progress,  both formal and practical, on a problem which is central to modern relativistic astrophysics.
The framework we have developed is ready to be applied and we expect to report progress on a set of relevant problems in the near future.

\acknowledgments

NA and IH gratefully acknowledge support from the STFC.

\appendix\markboth{Appendix}{Appendix}
\renewcommand{\thesection}{\Alph{section}}
\numberwithin{equation}{section}
\section{Vector Potential ``Entrainment''}
\label{appen}

 It may be worth noting that the Coulomb action in Eq.~\eqref{coulact} depends on 
the metric, and the combination $n^a_\x A_a$, which is exactly of the entrainment form, if we 
consider it as part of the fluid action $S_M$ (\cf Eq.~\eqref{varlamb}). \ This means that, at least formally, we can 
consider $\Lambda$ to be a functional of the set $\{n^2_\x,n^2_{\x \y},n^a_\x A_a\}$. \ This may be interesting as there are general constraints that can be had for plasmas if we 
make statements about gauge-invariance of the total fluid/plasma action and the vector 
potential entrainment. \ It is plausible that more general forms for the entrainment could lead 
to known results in, say, non-linear media in a more efficient way.

It is straightforward to work through the steps of varying the 
new action and obtaining the equations of motion:
\beq
       S = S_M\left(n^2_\x,n^2_{\x \y},A^2_\x\right) + S_{Max}\left(A_a\right) \ ,
\eeq
where
\beq
       A^2_\x \equiv - n^a_\x A_a \ .
\eeq
Next,
\bea
 \delta S &=& \fint \left[\sum_\x \mu^\x_a \delta n^a_\x - \sum_\x 
          \frac{\partial \Lambda}{\partial A^2_\x} \left(A_a \delta n^a_\x + n^a_\x \delta A_a\right) 
          \right. \cr
    && \left. + \frac{1}{2} \left(\Lambda g^{a b} + \sum_\x n^a_\x \mu^b_\x\right) 
          \delta g_{a b}\right] + \frac{1}{4 \pi} \fint \left(\nabla_a F^{a b}\right) \delta A_b \cr    
   && - \frac{1}{32 \pi} \fint \left(F_{c d} F^{c d} g^{a b} - 4 F^{a c} F^b{}_c\right) \delta g_{a b}
          \cr
 &=& \fint \left\{- \sum_\x \left(f^\x_a + \Gx \tilde{\mu}^\x_a - \rtotxa\right) \xi^a_\x - 
          \frac{1}{4 \pi} \left(\nabla_b F^{a b} - 4 \pi \sum_\x \mathcal{Q}_\x n^a_\x\right) 
          \delta A_a \right. \cr 
   && \left. + \frac{1}{2} \left[\Psi g^{a b} + \sum_\x n^a_\x \mu^b_\x - \frac{1}{16 \pi} 
          \left(F_{c d} F^{c d} g^{a b} - 4 F^{a c} F^b{}_c\right)\right] \delta g_{a b}\right\} \ , 
\eea
where
\beq
 \tilde{\mu}^\x_a = \mu^\x_a + \mathcal{Q}_\x A_a
                                  \quad , \quad
 \mathcal{Q}_\x \left(n^2_\x,n^2_{\x \y},A^2_\x\right) \equiv - 
                                  \frac{\partial \Lambda}{\partial A^2_\x} \ .
\eeq
We recover the minimal coupling when
\beq
       {\cal Q}_\x = e_\x \ .
\eeq

What happens if we now impose gauge-invariance on the whole system? \ We consider a 
variation of only the vector potential which is of the form
\beq
       \delta A_a = \nabla_a \delta \phi \ .
\eeq
Taking into account the identity in Eq.~\eqref{chcons}, we see that 
\beq
       \sum_\x \nabla_a \left({\cal Q}_\x n^a_\x\right) = 0 \ .
\eeq
This can also be written as
\beq
     \sum_\x \left({\cal Q}_\x \Gx + n_\x \frac{{\rm d} {\cal Q}_\x}{{\rm d} \tau_\x}
                           \right) = 0 \ .
\eeq
If ${\cal Q}_\x$ depends on only $X^A_\x$ (it is Lie-dragged by $u^a_\x$) then this reduces 
to
\beq
        \sum_\x {\cal Q}_\x \Gx = 0 \ .
\eeq

\end{document}